\documentclass[11pt,preprint]{aastex}

\slugcomment{Accepted to \aj}  
\shorttitle{LSB Galaxies and the TF relation} \shortauthors{Chung et al.}
\begin{document}

\title{\sc LSB Galaxies and the Tully-Fisher Relation}

\author{Aeree Chung} 
\affil{Department of Astronomy, Columbia University, New York, NY 10027}
\email{archung@astro.columbia.edu}

\bigskip

\author{J. H. van Gorkom}
\affil{Department of Astronomy, Columbia University, New York, NY 10027}
\email{jvangork@astro.columbia.edu}

\bigskip
\author{K. O'Neil}
\affil{Arecibo Observatory, HC03 Box 53995, Arecibo, PR 00612}
\email{koneil@naic.edu}

\and 
\author{G. D. Bothun}
\affil{Department of Physics, University of Oregon, Eugene, Oregon 97403}
\email{nuts@bigmoo.uoregon.edu}

\begin{abstract} 
We present Very Large Array (VLA) {\rm\sc Hi} imaging of four low surface 
brightness (LSB) galaxies ([OBC97] P01-3, [OBC97] C06-1, [OBC97] C04-1, 
[OBC97] C04-2)\footnote{The official (IAU) name of these objects are 
originated from \citet{one97a}. For convenience, however, [OBC97] will be 
omitted hereafter.} which were thought to strongly deviate from the 
Tully-Fisher (TF) relation based on Arecibo single-dish observations. We 
do not detect three of the four targeted LSB galaxies in {\rm\sc Hi} down 
to a 4$\sigma$ limit of 0.08Jy km s$^{-1}$. We find that two of the four 
of these LSBs have bright galaxies at projected distances of 2.6 and 2.9 
arcminutes, which have contaminated the Arecibo signal. A further 
examination of the Arecibo sample shows that five out of the six galaxies 
that were found to deviate from TF have nearby bright galaxies (within 
3.5 arcmin and within the observed {\rm\sc Hi} velocity range), and we 
conclude that possibly all but one of the non-TF galaxies are contaminated 
by {\rm\sc Hi} from nearby galaxies. The sixth galaxy was not detected by 
us. A new observation by Arecibo did not confirm the earlier detection. We 
present the {\rm\sc Hi} properties, kinematics, and rotation curves of 
three bright galaxies NGC 7631, KUG 2318+078, NGC 2595 which happened to 
be in the LSB fields and of two LSB galaxies, one of the targeted ones 
C04-2 and one (UGC 12535 or P01-1) that was found in the field of P01-3. 
The two detected LSB galaxies fall within 2$\sigma$ on the TF relation. 
The integrated profiles of the bright galaxies are consistent with the 
Arecibo results both in velocity range and amplitude, which indicates 
that most of the extreme deviators from the TF relation must have been 
affected by bright companions in this earlier {\rm\sc Hi} survey. A more 
recent determination of the sidelobe structure of the Arecibo beam 
supports our conclusion and shows that the degree of sidelobe 
contamination was much larger than could have been initially predicted. 
If the now unknown velocities of these non detected LSB's are within our 
probed velocity range then the limit to their {\rm\sc Hi} mass is roughly 
10$^8$ $M_\odot$ assuming $H_0=75$ km s$^{-1}$ Mpc$^{-1}$.  Hence, we have 
corrected the results found in \citet{one00a} and have reconstructed the 
TF relation. These new observations then show a) a reconstructed TF 
relation that has relatively large scatter at all values of rotational 
velocity (possibly indicating the true range in disk galaxy properties) 
and b) the presence of at least some red, gas poor, LSB disks that indeed 
may be in an advanced evolutionary state as the faded remnants of their 
former high surface brightness actively star forming state.
\end{abstract}


\clearpage
\newpage
\section{INTRODUCTION}

As sensitivity has improved, many galaxies fainter than the natural 
sky brightness have been discovered. It has now become clear that the 
central disk surface brightness of spiral galaxies does not fall within 
a narrow range of $\mu_B$(0)=21.65$\pm$0.30 mag arcsec$^{-2}$ as 
originally proposed by \citet{fre70} and that, at a given circular 
velocity, there is a very wide range of observed $\mu_B$(0). As a 
candidate repository of abundant baryonic mass in the Universe and a 
significant link in the unsolved evolutionary paths of galaxies, low 
surface brightness galaxies (LSB, hereafter) have  become a relatively 
new subject of study in extragalactic astronomy (Bothun et al. 1997 and 
references therein).

Since the distribution in surface brightness is continuous, there is no 
obvious definition for a low surface brightness galaxy, but, practically, 
many surveys define a low surface brightness system as a galaxy with a 
disk central surface brightness $\mu_B(0)\geq23$ mag arcsec$^{-2}$, a value 
roughly equal to the brightness of the night sky background in the regime 
between 4000 and 5000 \AA ~on a moonless night at a good astronomical 
observing site \citep{bot98}. Although several catalogs contained 
significant numbers of diffuse galaxies before the 1970s, the LSBs in those 
catalogs were limited to low mass galaxies and were not representative of the 
full range of LSB types. The discovery of LSB galaxies advanced considerably 
in the 1980s \citep{iab97}. To date (O'Neil et al. 1998 and references 
therein), three main classes of LSB galaxies have been identified: (1) dwarfs, 
defined by objects with scale lengths $\leq$1 kpc; (2) disk galaxies with 
scale lengths 1$\leq\alpha\leq$5 kpc and circular velocities in the range 
80-200 km s$^{-1}$; and (3) giant disk galaxies with scale lengths $\geq$5 
kpc. 

Dwarf LSBs include both irregular (dI) and more regular spiral (dS) 
galaxies \citep{sun98,sch01}. Dwarf LSB galaxies have much in common 
with blue compact dwarfs (BCDs): a large amount of {\rm\sc Hi}, often 
with small OB associations, and blue colors ($B-V\sim$0.5 mag). But dwarf 
irregulars are distinguished from BCDs by having amorphous shapes, as well 
as by having star-forming regions that are not centrally concentrated. Also, 
dwarf LSBs are known to have higher total masses (within a given surface 
brightness) and larger linear diameters than BCDs \citep{sun98}. Meanwhile, 
normal-size or giant LSBs are more similar to late-type spiral galaxies in 
their general properties, except in overall low stellar density. They have 
normal or rich {\rm\sc Hi} contents with $M_{\rm HI}$/$L_{B}$$\approx$1.5 
\citep{sch90, van93, deb96, pic97} but the average {\rm\sc Hi} surface 
densities are 6$\times$10$^{20}$ cm$^{-2}$ which is below the threshold for 
star formation \citep{van93, deb96}; CO emission is seldom detected in these 
systems \citep{sch90,one00b}. In a recent work by \citet{one02}, only 6 out 
of 28 LSB galaxies observed at mm wavelength have been detected in CO. The 
integrated $B-V$ colors are known to be unusually blue without large, bright 
star-forming regions \citep{sch90, mcg95, deb95} but giant LSB spirals are 
redder than smaller LSB galaxies with a trend of redder color with increasing 
\lq\lq{diffuseness}\rq\rq \citep{spr95b}.

Even though it is believed that giant LSBs may have evolutionary histories 
distinct from other LSBs \citep{mat01}, most of the LSB galaxies are thought 
to be quiescent, unevolved and gas-rich systems \citep{sch90, mab94, mcg95, 
deb96, mat01}. To explain localized star formation without global star 
formation in the LSB galaxies, {\lq\lq}sporadic star formation episodes and 
a slow evolutionary rate scenario\rq\rq~has been suggested \citep{deb95, 
bot97}. If the LSB galaxies evolve much slower than their high surface 
brightness counterparts, these systems must form less stars in a Hubble 
time and will thus have a higher total mass-to-light ($M/L$) ratio. For this 
reason, many groups have been studying the $M/L$ ratio and the gas content of 
LSB galaxies to understand how the star formation history and evolutionary 
state of these systems differ from those of high surface brightness galaxies.

The luminosity-{\rm\sc Hi} linewidth plane, first constructed by \citet{taf77}, 
indirectly contains information about the variance in global $M/L$ for disk 
galaxies. In simple dynamical terms, a small scatter about the TF relation can 
be understood if disk galaxies a) obey the virial theorem, b) have constant 
$M/L$ and c) have constant  $\mu_B$(0).   The mystery, of course, in the TF 
relation is that conditions b and c are demonstrably false.  The recognition 
that the potentials of disk galaxies are dark matter dominated makes the 
existence of the TF relation very difficult to understand from first principles 
because, apparently, it requires that the amount of luminosity trace the dark 
matter in a way that scales exactly with $V_c$. This fine tuning requires that 
the baryonic mass fraction in disk galaxy potentials be nearly constant, which 
should be a troubling point as it implies that just enough baryonic gas creeps 
into these dark potentials and then has the correct star formation history to 
produce the right amount of light for a given $V_c$.  In order for LSB disk 
galaxies to therefore comply with the TF relation requires an additional fine 
tuning between $\mu_B$(0) and $M/L$ must occur in such a way that higher $M/L$ 
potentials (e.g. those with a reduced baryonic mass fraction) still produce 
the correct amount of light for a given $V_c$.  
  
This apparent physical paradox strongly suggests that LSB galaxies which 
deviate from the observed TF relation must exist. Thus, searches have been 
conducted for any strong discrepancies from the TF relation among LSB 
galaxies in order to see whether or not there exists such a conspiracy, 
and how it could affect the star formation histories of either LSBs or 
normal spiral galaxies. Recently, \citet{one00a} found six apparently 
extreme systems in their {\rm\sc Hi} survey using Arecibo, which appear to 
strongly deviate from the TF relation thus opening up the manifold of disk 
galaxy properties.  This potentially important result  was highlighted
by \citet{one00a}, but, unfortunately, as demonstrated in this paper, that
result is an artifact of a much larger than anticpated side-lobe problem 
associated with the reconfiguration of the Arecibo focussing system and the 
subsequent change in illumination pattern on the dish.

The sample of \citet{one00a} contains a more diverse LSB population in terms 
of color and size than other samples. In 1997, O'Neil, Bothun and Cornell 
\citep{one97a, one97b} conducted a comprehensive survey of LSB galaxies in 
order to obtain structural parameters such as $\mu_{B}$, $\alpha$, $r_{25}$, 
etc. and colors (Johnson/Cousins ${U,~B,~V,~I,~\&~R}$, when possible) in the 
Cancer and Pegasus galaxy clusters (hearafter OBC) using the University of 
Texas MacDonald Observatory 0.8 m telescope. In total 127 galaxies were found 
with $\mu_{B}\geqslant$22.0 mag arcsec$^{-2}$ including 119 newly identified
LSB systems. The colors of these galaxies range continuously from very blue 
($U-B$=0.56, $B-V$=0.37) to very red ($B-V>$0.8; note that galaxies of S0a or 
earlier types have this $B-V$ value in general - see Figure 5 of \citet{rah94}), 
which indicates that LSB galaxies at the present epoch define a wide range of 
evolutionary states \citep{one97b}. Also, the OBC catalog includes many small 
galaxies with scale lengths less than 1 kpc. Since the redshifts were not 
known, the conversion from arcsec to kpc was based on an average distance 
either to the cluster (Pegasus or Cancer) or to the primary galaxy in the 
image \citep{one97a}. 

More recently, \citet{one00a} conducted an {\rm\sc Hi} survey of galaxies 
from the OBC catalog using the refurbished 305m Arecibo Gregorian Telescope, 
and detected {\rm\sc Hi} emission from 43 galaxies out of the 111 LSBs in 
the 2000-12000 km s$^{-1}$ range. The detected galaxies cover a similar range 
of properties as HSB galaxies, ranging from very blue through very red 
($B-V$=-0.7 - 1.7) and from extremely gas rich through gas poor ($M_{\rm 
HI}$/$L_{B}$=0.1 - 50 $(M/L)_{\odot}$). With the determined line widths and 
absolute magnitudes, the authors find that only 40\% of the {\rm\sc Hi} 
detected LSB galaxies fall within 1$\sigma$ of the standard high surface 
brightness galaxy TF relation. In addition, they find six extreme outliers 
(see Fig. 1). This indicates that the $M/L$ ratios of some LSB galaxies are 
much higher than the values needed to compensate for low surface brightness. 
In order to determine the $M/L$ more accurately and to study the detailed 
{\rm\sc Hi} properties such as the {\rm\sc Hi} distribution, the isophotal 
({\rm\sc Hi}) size, the local kinematics, rotation curves, and $M_{\rm HI}
$/$M_{T}$, it is required to get  {\rm\sc Hi} synthesis images and in this 
paper we investigate the galaxies most discrepant from the TF relation found 
by \citet{one00a}. We obtained images with the VLA of four of the 43 galaxies 
detected in {\rm\sc Hi} by \citet{one00a} - three out of the six extreme 
non-TF galaxies and one galaxy which deviates from the TF relation by 2$\sigma$, 
using the VLA in the 3 km C-short configuration. The observations in the VLA, 
however, show that some of the Arecibo results are contaminated by bright 
galaxies, due to a significantly larger sidelobe compared to what was 
believed at the time of observation.


In this paper, we present the results of the VLA observations. In \S~\ref{obs}, 
we summarize the observations and data reduction. In \S~\ref{result}, we present 
and discuss the {\rm\sc Hi} properties of the galaxies which we have found in 
the field. Finally, in \S~\ref{discuss}, we compare the {\rm\sc Hi} global 
profiles of the bright galaxies (contaminators) to the {\rm\sc Hi} spectra 
of the LSBs obtained by \citet{one00a} to illustrate the degree of confusion 
in the Arecibo {\rm\sc Hi} survey caused by the extended sidelobe.  In light 
of this, this paper then corrects the results of \citet{one00a} and removes 
some of the observed TF discrepancy thus maintaining the mystery of how it is 
that LSB disk galaxies can populate a similar TF locus as high surface 
brightness galaxies.

\section{OBSERVATIONS AND DATA REDUCTION}
\label{obs}

The {\rm\sc Hi} imaging was done in June 2000 with the VLA in 
the 3 km C-short configuration. We observed three out of the six 
extreme departures from the TF (P01-3, C06-1, C04-1) and C04-2 which is 
located at the 2$\sigma$ upper limit of the TF relation (see Fig. 1). 
The 21-cm linewidths (at the 50\% level) determined by \citet{one00a} 
for these sample galaxies are 375, 333, 410 and 85 km s$^{-1}$, with peak 
flux densities of 9, 8, 9 and 6 mJy \citep{one00a}. From the known velocity 
widths, the bandwidth was chosen as 6.25 (for P01-3 and C06-1) and 3.125 (for 
C04-1 and C04-2) MHz. The data were obtained with 2 polarizations (2AC correlator 
mode) and 63 channels. The field of view is set by the 25 m diameter of the 
individual array elements. We integrated 8 hours per galaxy resulting in a 
2$\sigma$ column density sensitivity of 5.6 and 4.8$\times$10$^{19}$ cm$^{-2}$ 
per channel ($\Delta v$=20.6 km s$^{-1}$) for P01-3 and C06-1, respectively, 2.6 
and 3.2$\times$10$^{19}$ cm$^{-2}$ per channel ($\Delta v$=10.3 km s$^{-1}$) 
for C04-2 and C04-2, respectively. Instrumental parameters and resulting 
sensitivities are listed in Table 1. \footnote{Note that the rms per 
channel of P01-3 cube is worse or comparible to the cubes of C04-2 and 
C04-1 even though its wider channel width and this is caused by a 
serious solar interference on that day.}


Standard VLA calibration procedures were applied in AIPS. 
The continuum was subtracted by making a linear fit to the 
visibility data for a range of line-free channels in both sides 
of the band. We chose line-free channels based on the {\rm\sc Hi} 
spectra of \citet{one00a}. Two of the observations (P01-3 and C04-1) 
were seriously affected by solar interference. For these we first 
subtracted a model of the continuum image from the $uv$ data, and 
then clipped $uv$ data above 2 Jy for each datacube. To make the images
we applied a weighting scheme intermediate between uniform and natural but 
closer to a natural weighting scheme to maximize sensitivity. 

\section{RESULT}
\label{result}

In our VLA observations of the LSB galaxies, we detect five 
galaxies in total (including two LSBs: UGC 12535 (P01-1) and C04-2, 
from the four {\rm\sc Hi} datacubes centered on the LSB galaxies. In 
the field of P01-3, we detect three galaxies, -NGC 7631, KUG 2318+078 
and UGC 12535-but not P01-3. UGC 12535 is in fact the same as the LSB 
galaxy P01-1 in the OBC catalog. Likewise, in the field of C06-1, NGC 2595 
was detected instead of the LSB galaxy, C06-1. In fact, the {\rm\sc Hi} 
profiles of P01-2, P01-3 and C06-1 of \citet{one00a} are consistent with those 
of KUG 2318+078, NGC 7631 and NGC 2595, respectively which implies that the 
{\rm\sc Hi} survey of \citet{one00a} must have been contaminated by bright 
neighbors. The details will be discussed in \S~\ref{discuss}. We do not 
see any {\rm\sc Hi} from C04-1 which appears to be the only isolated LSB 
deviating significantly from the TF relation (Fig.1 and Fig. 2). 


Below, we present the {\rm\sc Hi} properties and {\rm\sc Hi} 
maps of the galaxies which we detect from the VLA run (Jun, 2000). 
The extended {\rm\sc Hi} properties are summarized in Table 2. The 
{\rm\sc Hi} mass is calculated using $M_{\rm HI}=2.36\times 10^{5}d^{2}{\int} 
S_{\rm HI} dv~M_{\odot}$, where the distance $d$ in Mpc and the integrated 
{\rm\sc Hi} emissivity ${\int}S_{\rm HI} dv$ is in Jy km s$^{-1}$. We assume 
$H_0=75$ km s$^{-1}$ Mpc$^{-1}$ in this work. The {\rm\sc Hi} linewidths, $W_{20}$ 
and $W_{50}$, were measured in three different ways: at 20\% and 50\% of (1) the 
highest peak; (2) two different peaks of each side of the systemic velocity; (3) 
the average of those two peaks. The differences are not larger than 10 km s$^{-1}$ 
and the linewidths determined by the second method are given in the Table. The 
smallest {\rm\sc Hi} mass that can be detected in a datacube at the position of 
the LSB is a 4$\sigma$ signal per beam per channel. The upper limit to the {\rm\sc 
Hi} mass is $\sim 10^8 $M$_\odot$, if the LSB's are within the velocity range 
probed. Since we now no longer have a detection of these galaxies, we do not 
know their redshifts and they may lie outside the observed velocity range. 


\subsection{LSB galaxies}

We detect {\rm\sc Hi} from two LSB galaxies, 
UGC 12535 (Fig. 3) and C04-2 (Fig. 4). 

UGC 12535 is found in the field of P01-3 and is also known as
P01-1 in the OBC catalog. UGC 12535 was also in the original 
cluster spiral sample of \citet{bot82} who failed to detect it in 21-cm
and noted that it had an ususually low value of $M_{HI}/L_B$ for its classification
as a normal-sized late-type LSB galaxy with $D_{25}\approx$17.5 kpc (assuming $H_0$=75 
km s$^{-1}$ Mpc$^{-1}$).  These new observations show that UGC 12535 is less 
than 2$\sigma$ from TF relation with the corrected {\rm\sc Hi} linewith, $W_{50}^
{\rm Corr}$=174 km s$^{-1}$ (see Fig. 11). Our {\rm\sc Hi} mass of UGC 12535
implies an $M_{HI}/L_B$ of 0.15 which is indeed very low. This is a relatively 
red LSB disk with 
observed color of $B-V$ = 0.82 (see both O'Neil et al. 1997a and Bothun et al. 1985)
and thus represents a rare case of a LSB disk with relatively low gas content
which may indeed be the faded evolutionary remains of a former HSB spiral, of
which there are currently many in the Pegasus I cluster (see \citet{bot82}).

Another {\rm\sc Hi} detected LSB galaxy is C04-2. C04-2 is one of the 
newly identified LSBs in {\citet{one97a}}'s CCD survey for LSB galaxies. 
However, in \citet{one97a}, the scale length of C04-2 is not available since 
they did not provide information for galaxies with King profiles or galaxies 
which have such irregular surface brightness profiles that they are unsuitable 
for a linear fit. But considering its apparent size (D$_{25}=22.1''$) and radial 
velocity ($V_{opt}=5168$ km s$^{-1}$), the physical size (diameter) of C04-2 is about 
7.4 kpc and it must be either of the first or second class. The {\rm\sc Hi} profile 
of C04-2 is consistent with the {\rm\sc Hi} spectrum of \citet{one00a}. Since the 
size of C04-2 is comparable to the beamsize itself, we could not derive the 
rotation. C04-2 is also one of the very red LSBs with $B-V$=1.33 and 
deviates from the TF by 2$\sigma$ in the direction of smaller $M/L$.


The star formation history of LSB disk galaxies has generally been hard to
understand, mostly because of the bulk of these disk galaxies are very blue
but with very low current star formation rates. The simplest traditional scenario, 
that LSB disks are the faded remnants of HSB disks and thus represent an advanced 
state of disk galaxy evolution, has long been ruled out by the fact that the LSB 
disk is very blue and gas rich in general. This study has now produced a good 
candidate in UGC 12535 as a faded HSB disk. (In addition, the decontaminated
red LSBs now remain as such with no known velocity.  If they are indeed cluster
members, as seems likely given their angular size, then they too may comprise
additional examples of red LSB disks that are gas poor).  On the other hand, C04-2
is very red yet has relatively high gas content which straightforwardly shows
that  red LSB galaxies are not necessarily gas-poor, a result consistent with
\citet{one00a}.   Thus, like their blue counterparts, red LSB disks are unlikely
to be in the same evolutionary state at the present time.

\subsection{HSB neighbors}

NGC 7631 (Fig. 5) is found in the field of P01-3 with a radial velocity 
of 3746 km s$^{-1}$. We see its rotation curve declining (on both sides). 
The rotation velocity of 200 km s$^{-1}$ decreases between 17 kpc and 20 kpc, 
beyond R$_{25}$ ($\approx$13.7 kpc), by 15$\pm$5.7 km s$^{-1}$ ($\sim$7\% of 
the maximum rotation velocity). Previously, \citet{cav91} studied spiral galaxies 
which have declining rotation curves and find falling curves are more dominant in 
bright, compact galaxies and rising curves are more frequently found in low-luminosity
galaxies. They conclude that the decrease in rotational velocity indicates a large 
ratio of luminous to dark mass in luminous regions of the systems, which suggests 
that the compactness of a galaxy is a measure of the initial spin parameter. We 
see a declining rotation curve from another bright, compact galaxy, KUG 2318+078 
(Fig. 6), found in the field of P01-3 with 3886 km s$^{-1}$. KUG 2318+078 has 
remarkably asymmetric kinematics and the decrease of the rotational velocity is 
found only in one side. In the approaching side, the rotational velocity reaches 
at its maximum velocity, 125.0$\pm$1.8 km s$^{-1}$ at a radius between 7 and 8 kpc, 
then drops by $\sim$21\% to 98.6$\pm$6.0 km sec$^{-1}$. The {\rm\sc Hi} profile is 
consistent with the {\rm\sc Hi} spectrum of P01-2 (Fig. 8) obtained at Arecibo. KUG 
2318+078 seems to be located outside of the Arecibo main beam (Fig. 9), but it 
must have affected P01-2 within the first sidelobe. The 
details will be discussed in the last section. 

In the field of C06-1, NGC 2595 (Fig. 7) is found at a distance of 
2.9 arcmin from C06-1 with a radial velocity of 4214 km s$^{-1}$. NGC 
2595 has well defined spiral arms and a flat rotation curve with the 
maximum velocity of 325$\pm$17 km s$^{-1}$.  


NGC 7631 and NGC 2595 fall into the middle of the standard TF, while 
KUG 2318+078 is located between 1$\sigma$ and 2$\sigma$ to the direction 
of over luminous for a given linewidth (Fig. 11).  


\section{DISCUSSION}
\label{discuss}

Through VLA synthesis imaging (2000), we fail to detect 
{\rm\sc Hi} from any of the LSB galaxies which were thought 
by \citet{one00a} to deviate significantly from the TF relation. 
Instead, we find that the {\rm\sc Hi} profiles of NGC 7631 and NGC 
2595 are consistent in velocity range and amplitude with P01-3 and C06-1 
observed by \citet{one00a} using the Arecibo telescope (using the corrections 
found by \citet{hei01}). The bright galaxies, NGC 7631 and NGC 2595 are located 
at 2.6 and 2.9 arcmin from P01-3 and C06-1, respectively. Therefore, we conclude that 
the {\rm\sc Hi} profiles of P01-3 and C06-1 must have been contaminated by those bright 
companions. Similarly, we find that the {\rm\sc Hi} profile of P01-2 of \citet{one00a} 
is consistent with that of KUG 2318+078 which is 2.9 arcminute away (see Fig. 8). In 
none of these cases, the bright contaminators were expected to seriously affect the 
signal of the LSB galaxies in the center considering the beamsize of the Arecibo at 
1420 MHz (3.1$\times$3.5 in Az$\times$ZA). However, recent work by \citet{hei01} 
shows that the first sidelobe extends to $\sim$ 10 arcmin diameter and if we 
take this into account, the {\rm\sc Hi} survey by \citet{one00a} could have 
been seriously contaminated.  As an attempt to save face, we point out that,
at the time of the original observations (scheduled to be the first extragalactic
post upgrade Arecibo observations) \citet{one00a} would not have even undertaken
these observations had they known that the post upgrade beam now contains a sidelobe
which is effectively 3 times the beam size!  Clearly that produced some anamolous
results, which scientifically were interesting and lead to the VLA follow-up
observations which then showed how severe the contamination really was.

Indeed, five out of the six LSB galaxies which were thought to 
deviate significantly from the TF relation \citep{one00a} have bright 
companions within the first sidelobe of Arecibo (Fig. 10). Besides P01-3, 
C06-1, and P01-2, two other extreme non-TF LSB galaxies, C05-5 and C08-3, must 
have also been affected by the bright, nearby galaxies UGC 4416 and UGC 4308. 
In Table 3, we present the list of LSB galaxies and their possible contaminators 
with coordinates and radial velocities. Note, however, that unlike most of the LSB 
galaxies which have bright neighbors, P01-4 does not seem to be contaminated, even 
though P01-4 has a bright neighbor, NGC 7631, at a distance of 3.4 arcmin and it 
falls within the distance the beam to the sidelobe. It implies that the first 
sidelobe might be asymmetric, consistent with the results of \citet{hei01}. We 
investigated the fields of the 43 {\rm\sc Hi} detected LSB galaxies in \citet
{one00a}'s sample. Excluding the five galaxies which turned out to be 
contaminated by bright galaxies, the rest of the {\rm\sc Hi}-detected LSB 
galaxies are safely isolated within the first sidelobe of the Arecibo. 


From this work, we conclude that we do not see any evidence of strong 
deviation from the TF relation for the LSB sample of \citep{one00a}. Even 
though many groups have been working on this subject, there is still no agreement 
if there exists any fine tuning between central surface brightness and the total 
mass-to-light ratio that makes disk galaxies follow the TF relation. For instance, 
\citet{zwa95} and \citet {spr95a} find that the LSB galaxies follow the same TF 
relation defined by normal spiral while \citet{pas91} and \citet{mat98b} find a 
curvature (underluminous at a given line width) at the low luminosity end of 
the TF for LSB galaxies. 

These contradictory results may partly be due to selection 
effects. As \citet{mat98b} point out, the star formation history 
and evolutionary states of LSB galaxies studied using the TF relation 
gives different interpretations strongly depending on the sample galaxies. 
For example, investigators who studied comparable to or larger than normal-sized 
galaxies conclude that LSB galaxies do follow the same TF relation \citep{zwa95, 
spr95a}, while others who included smaller LSBs find larger scatter 
in the TF \citep{mat98b}. Here we have a good sample (OBC catalog, 1997) 
which is diverse in terms of sizes and colors and in fact, we see a 
larger scatter even if we exclude the contaminated LSBs.


In Fig. 11, we present the TF relation of \citet{one00a}'s LSBs 
excluding the contaminated LSBs at $B$-band. Indeed $B$-band TF
relations are often known to suffer from star formation activities 
or internal extinction. However, the ratio of surface brightness of 
the galaxy to the sky background steadily decreases as one goes to longer 
wavelength. Therefore, the S/N of low surface brightness systems at longer
wavelength such as $I$-band is lower than at $B$-band, which still makes us 
rely on $B$-band TF relation the most rather than TF relations at any
other longer wavelengths for low surface brightness galaxies (see Fig. 12). 
Hence, here we discuss about the TF relation of \citet{one00a}'s LSBs only 
at $B$-band through Fig. 11.

We see a larger scatter at the 
low luminosity end but this is typical of most TF samples (see also
\citet{mcg01}). We find, $M_B^{Corr}$=-(6.57$\pm$0.96)log$W_{50}^{Corr}$-3.84$\pm$1.91
with 38 LSB galaxies which is close to the TF relation obtained by \citet{zwa95} 
($M_B^{Corr}$=-6.59log$W_{50}^{Corr}$-3.73$\pm$0.77 from 42 sample galaxies) but
gives a larger scatter.
Much of this scatter at the low luminosity end
can be attributed to inclination error and/or significant non-circular
motions that contributed to the observed linewidth.  What is therefore
more relevant is that we see significant scatter at higher linewidths
(e.g. $>$ 100 km s$^{-1}$) and this is almost surely reflecting the intrinsic
variation in global $M/L$ among disk galaxies of varying $\mu_B$(0).

In this figure, LSBs are also plotted by color; 
very blue ($B-V<$0.6), intermediate color (0.6$\leq B-V \leq$0.8) and 
very red ($B-V>$0.8). Here we find a weak trend of scatter by colors; 
blue LSBs seem to have higher $M/L$ (as \citet{zwa95} or \citet{mat98a} 
find), while no preferential direction is seen for red LSBs in their scatter. 
In fact, the OBC catalog contains many red small LSBs compared to previous 
LSB studies and from some of the red LSB samples, we find an opposite 
trend of the curvature which \citet{pas91} or \citet{mat98b} found. 
However, the integrated color in this plot is drawn using $B-V$ 
only and it requires further study.

We learn the following from the TF of the LSB galaxies in the OBC 
catalog \citep{one00a}: (1) the curvature predicted and found by \citet
{pas91} \& \citet{mat98b} is marginally found only for the blue LSB galaxies; 
(2) unlike blue ($B-V<0.4$) LSBs, very red ($B-V>0.8$) LSB galaxies do not show 
any preferential direction in deviating from the standard TF; (3) a weak correlation 
between gas richness and the deviation from the TF, is found in agreement with the 
findings by \citet{mat98b}.  This broadening of the TF relation, as found in
this study, is consistent with moving away from selection effect dominated
systems and towards a more representative sample of disk galaxies.

\acknowledgments
The VLA of the National Radio Astronomy Observatory 
is operated by Associated Universities, Inc. under a 
cooperative agreement with the National Science Foundation.
This work has been supported in part by NSF grant AST-00-98249 
to Columbia University.

\clearpage 
\begin{deluxetable}{lcccc} 
\tabletypesize{\scriptsize}
\tablecaption{VLA observing parameters. \label{tbl:tbl1}} 
\tablewidth{0pt} 
\tablehead{ & P01-3 & C06-1 & C04-2 & C04-1 } 
\startdata 
Phase Center: & & & & \\ 
R.A. (2000)\tablenotemark{a} & 23 21 18.2 & 08 27 31.5 & 08 23 29.3 & 08 24 33.1 \\ 
Dec. (2000)\tablenotemark{b} & 08 14 29.0 & 21 30 04.0 & 21 36 45.0 & 21 27 04.0 \\ 
Velocity Center (km s$^{-1}$)    & 3746 & 4322 &  5168 &  7905 \\ 
Velocity Range  (km s$^{-1}$)    & 1250 & 1250 &   630 &   630 \\
Time on Source (hrs)             &  7.8 &  7.8 &   7.8 &   7.7 \\ 
Bandwidth (MHz)                  & 6.25 & 6.25 & 3.125 & 3.125 \\ 
Number of Channels               &   63 &   63 &    63 &    63 \\ 
Channel Separation (km s$^{-1}$) & 20.6 & 20.6 &  10.3 &  10.3 \\ 
Synthesized Beam FWHM\tablenotemark{c}  & $19\times16$ & $17\times16$ 
                                 & $18\times16$ & $17\times16$ \\ 
Noise level (1$\sigma$):  & & & & \\ 
-rms noise (mJy beam$^{-1}$)     & 0.36 & 0.28 &  0.32 &  0.37 \\ 
-rms noise (10$^{19}$ cm$^{-2}$) &  2.8 &  2.4 &   1.3 &   1.6 \\ 
\enddata 
\tablenotetext{a}{R.A. in (${h~~~m~~~s}$)}
\tablenotetext{b}{Dec. in ($^{\circ}~~~~'~~~~''$)}
\tablenotetext{c}{in arcsec$^{2}$}
\end{deluxetable}

\clearpage 
\begin{deluxetable}{llccccc} 
\tabletypesize{\scriptsize}
\tablecaption{General and {\rm\sc Hi} properties of {\rm\sc Hi} detected galaxies. 
\label{tbl:tbl2}} 
\tablewidth{0pt} 
\tablehead{ 
& & \multicolumn{2}{c}{LSB galaxies} 
  & \multicolumn{3}{c}{HSB galaxies} \\
& & \multicolumn{2}{c}{-----------------------------} 
  & \multicolumn{3}{c}{------------------------------------------------------} \\
& units & UGC 12535\tablenotemark{a} & C04-2\tablenotemark{b} 
& NGC 7631\tablenotemark{a} & KUG 2318+078\tablenotemark{a} 
& NGC 2595\tablenotemark{a}} 
\startdata 
\multicolumn{2}{l} {General properties:}  & & & & & \\ 
\multicolumn{2}{l} {Position (2000.0) -}  & & & & & \\ 
R.A.\tablenotemark{c} & &~~23 21 01.6 &~~08 23 29.4 &~~23 21 26.7 &~~23 21 05.8 &~~08 27 42.0 \\
Dec.\tablenotemark{d} & & +08 10 45.9 & +21 36 48.0 & +08 13 03.0 & +08 10 45.9 & +21 28 44.0 \\ 
Type             &             &  Sbc &   Im &   Sb & SBbc &  Sbc \\ 
Distance         &         Mpc & 56.2 &68.9\tablenotemark{e}& 50.4 & 51.8 & 57.8 \\ 
$D_{25}$         &      arcmin & 1.01 & 0.37 & 1.82 & 1.02  & 2.57 \\ 
$i_{opt}$        &     degrees & 90.0 & 31.1 & 70.4 & 67.8  & 42.4 \\ 
$V_{opt}$        & km s$^{-1}$ & 4214 &5168\tablenotemark{c}& 3778 & 3886 & 4333 \\ 
$B_{\rm{T}}^{0}$ &         mag &14.80 &17.69 &13.04 &14.50  &12.81 \\ 
& & & & & & \\ 
\multicolumn{2} {l} {{\rm\sc Hi} properties:}  & & & & & \\ 
$S_{\rm HI}$ & Jy km s$^{-1}$ & 0.62$\pm$0.18 & 0.13$\pm$0.08 
                              &  4.5$\pm$0.20 &  2.6$\pm$0.18 & 19.9$\pm$0.17 \\ 
$V_{\rm HI}$ & km s$^{-1}$ &4216$\pm$7  &5168$\pm$6 &3747$\pm$10&3881$\pm$5 &4331$\pm$5 \\ 
$V_{rot}$    & km s$^{-1}$ & 163$\pm$15 &$\cdot$    & 180$\pm$5 &  55$\pm$2 & 325$\pm$10\\
$i_{\rm HI}$ & degrees     &  23$\pm$10 &$\cdot$    &  55$\pm$2 &  69$\pm$4 &  26$\pm$2 \\
$W_{20}$     & km s$^{-1}$ & 202$\pm$10 & 107$\pm$5 & 383$\pm$10& 162$\pm$10& 331$\pm$10\\
$W_{50}$     & km s$^{-1}$ & 189$\pm$10 &  96$\pm$5 & 360$\pm$10& 124$\pm$10& 304$\pm$10\\ 
$M_{\rm HI}$ &10$^{9}$ $M_{\odot}$  &   0.46$\pm$0.13 & 0.15$\pm$0.09 
                           & 2.69$\pm$0.12 & 1.65$\pm$0.11& 15.7$\pm$0.13 \\ 
\enddata

\tablenotetext{a}{Optical data referred to LEDA (Lyon-Meudon Extragalactic Database)}
\tablenotetext{b}{Optical data referred to \citet{one97a,one97b}}
\tablenotetext{c}{R.A. in (${h~~~m~~~s}$)}
\tablenotetext{d}{Dec. in ($^{\circ}~~~~'~~~~''$)}
\tablenotetext{e}{Determined by \citet{one00a} using the Arecibo}
\end{deluxetable}

\clearpage 
\begin{deluxetable}{ccccccccc} 
\tabletypesize{\scriptsize}
\tablecaption{LSB Galaxies classified as extreme non TF and possible contaminators. 
\label{tbl:tbl3}} 
\tablewidth{0pt} 
\tablehead{ 
\multicolumn{4}{c} {LSB Galaxies \citep{one00a}} & 
\multicolumn{4}{c} {Possible Contaminators\tablenotemark{a}} & Distance \\ 
\multicolumn{4}{c} {--------------------------------------------------------} & 
\multicolumn{4}{c} {~~~~-----------------------------------------------------------------} & between \\
Name & R.A. (2000) & Dec. (2000) & $V_{hel}$ & 
Name & R.A. (2000) & Dec. (2000) & $V_{hel}$ & two galaxies\\
& $h~~~m~~~s$ & $~~^{\circ}~~~{'}~~~{''}$ & km s$^{-1}$ & 
& $h~~~m~~~s$ & $~~^{\circ}~~~{'}~~~{''}$ & km s$^{-1}$ & (in arcmin) } 
\startdata 
C05-5 & 08 27 11.8 & +22 53 47 & 5524 &     UGC 4416 & 08 27 16.8 & +22 52 40 & 5533 & 1.67 \\ 
C06-1 & 08 27 31.5 & +21 30 04 & 4322 &     NGC 2595 & 08 27 42.0 & +21 28 44 & 4330 & 2.94 \\ 
C08-3 & 08 17 16.3 & +21 39 14 & 3570 &     UGC 4308 & 08 17 25.8 & +21 41 08 & 3566 & 3.04 \\ 
P01-2 & 23 21 16.1 & +08 04 54 & 3828 & KUG 2318+078 & 23 21 05.8 & +08 06 10 & 3886 & 2.87 \\ 
P01-3 & 23 21 18.2 & +08 14 29 & 3746 &     NGC 7631 & 23 21 26.7 & +08 13 03 & 3754 & 2.56 \\ 
P01-4 & 23 21 36.1 & +08 15 28 & 3890 &     NGC 7631 & 23 21 26.7 & +08 13 03 & 3754 & 3.37 \\ 
\enddata 
\tablenotetext{a}{Data from LEDA (Lyon-Meudon Extragalactic Database)}
\end{deluxetable}

\clearpage
\begin{figure}
\plotone{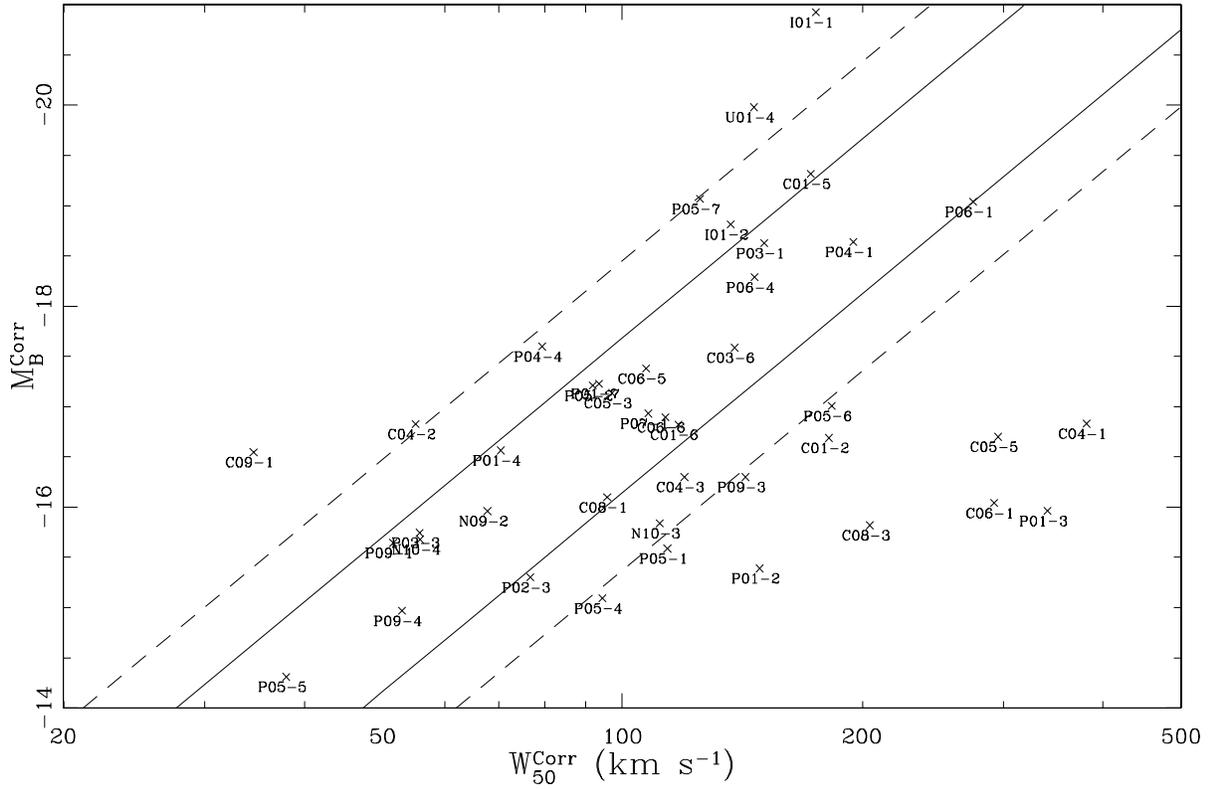}
\caption{\footnotesize
Absolute magnitudes vs. {\rm\sc Hi} linewidths of LSB galaxies determined 
by \citet{one00a}. The absolute magnitudes were not corrected for extinction. 
Solid lines and dashed lines correspond to 1$\sigma$ and 2$\sigma$ of the 
LSB galaxy Tully-Fisher relation (\citet{zwa95}), respectively. In this 
figure, we see P01-2 is quite far off from the Tully-Fisher relation instead of P09-4 
which falls into the right center of the TF relation but was classified 
as one of the extreme non TF cases by \citet{one00a}. Note, the six most 
deviant points are P01-2, P01-3, C04-1, C05-5, C06-1 and C08-3. \label{fig:fig1}}
\end{figure}

\clearpage
\begin{figure}
\plotone{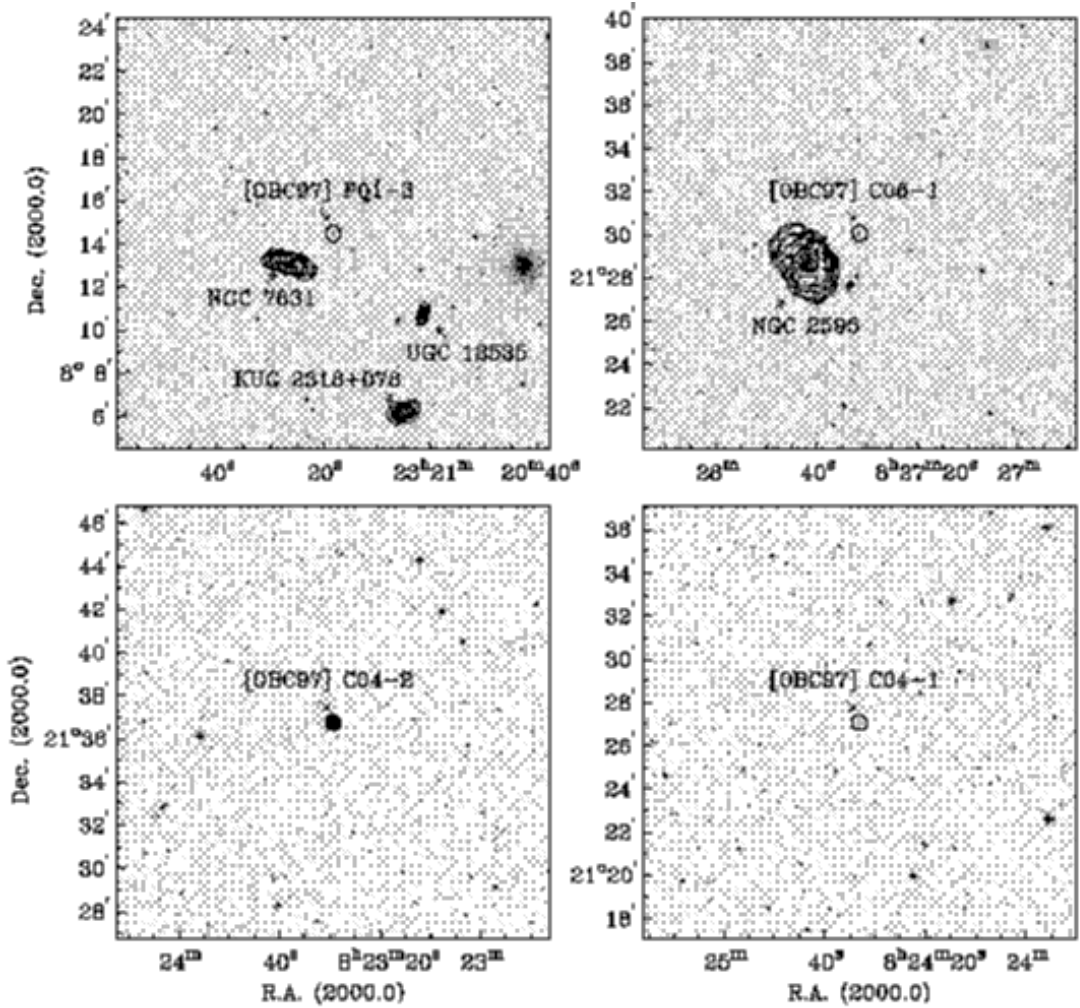}
\caption{\footnotesize
The HI contours overlaid on optical images ($20^{'}\times20^{'}$). 
The HI images were obtained using the VLA, June, 2000. Top left) NGC 7631, 
KUG 2318+078, UGC12535 are found in the field of P01-3, while P01-3 does not 
show any {\rm\sc Hi} signal. UGC 12535 is an identical object with P01-1 in 
OBC catalog. Top right) NGC 2595 is detected instead of C06-1. Bottom left) 
C04-2, one of the two HI detected LSB galaxies in the VLA run. Bottom right) 
We do not see any {\rm\sc Hi} from C04-1 in the VLA observation even though 
this galaxy is apparently the only isolated extreme non TF LSB among \citet
{one00a}'s sample galaxies and a recent Arecibo observation did not confirm
the previous result of \citet{one00a}.\label{fig:fig2}}
\end{figure}

\clearpage
\begin{figure}
\plotone{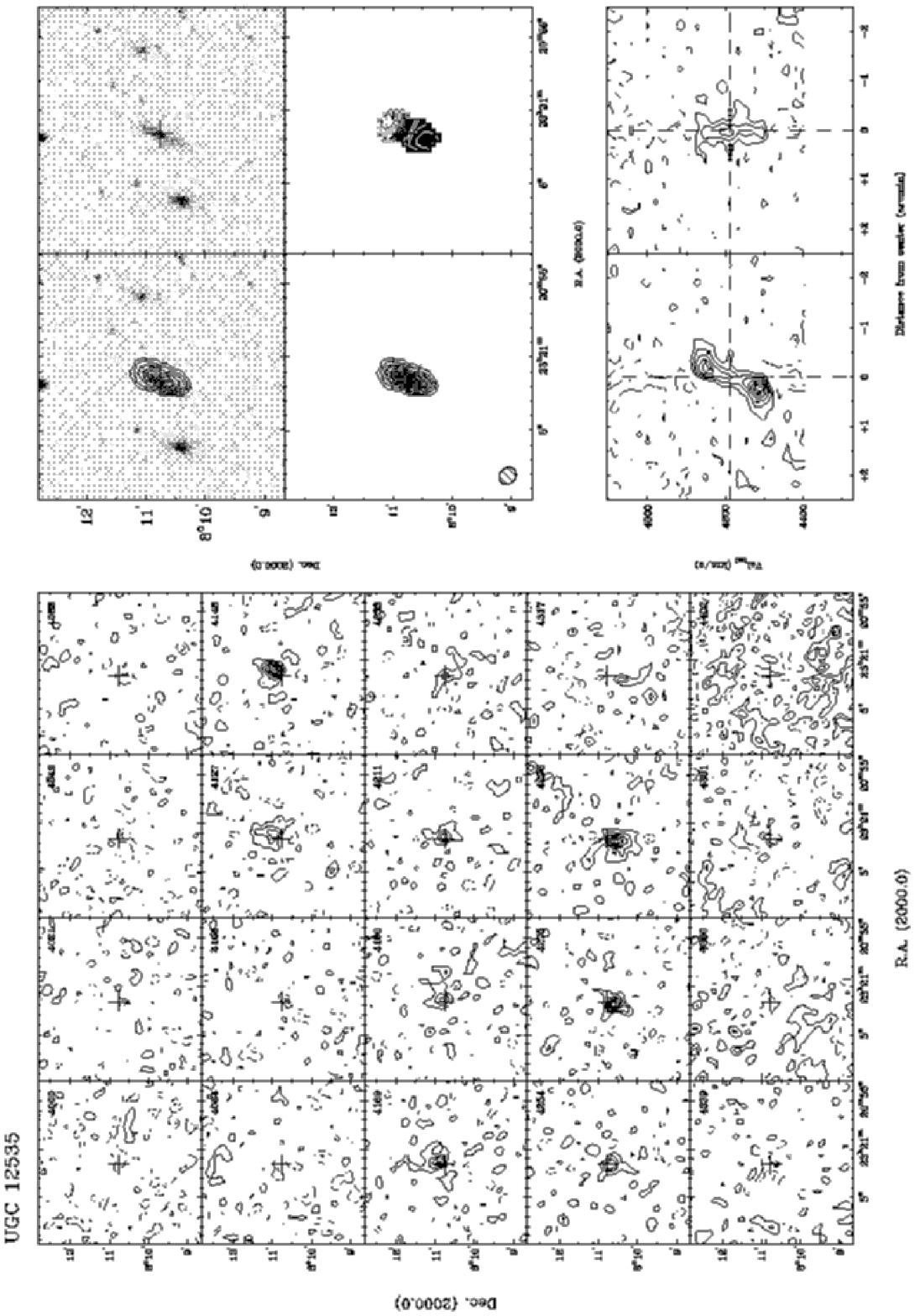}
\caption{\footnotesize
UGC 12535 ($4.25^{'}\times4.25^{'}$); LSB.
Left) Channel maps. The cross in each box indicates the optical center. 
The velocities in the upper right conners are heliocentric. 
Contour levels are 0.54$\times$1, 2, 3... mJy beam$^{-1}$ (solid lines) 
and -0.54$\times$1, 2 mJy beam$^{-1}$ (dashed lines). 
Right-Top) Counterclockwise from the top right, the optical image; 
the HI contours on the optical image (contours are 0.64+1.28$\times$1, 2, 3,... 
10$^{19}$ cm$^{-2}$ with the peak of 7.26$\times$10$^{19}$ cm$^{-2}$); 
the HI contours on the grayscales; the velocity field map (4216.4$\pm$15 km s$^{-1}$). 
Right-Bottom) Position-velocity cuts - along the major axis (left) 
and the minor axis (right). Contour levels are 0.53$\times$1, 2, 3... mJy beam$^{-1}$ 
(solid lines) and -0.53$\times$1, 2 mJy beam$^{-1}$ (dashed lines).\label{fig:fig3}}
\end{figure}

\clearpage
\begin{figure}
\plotone{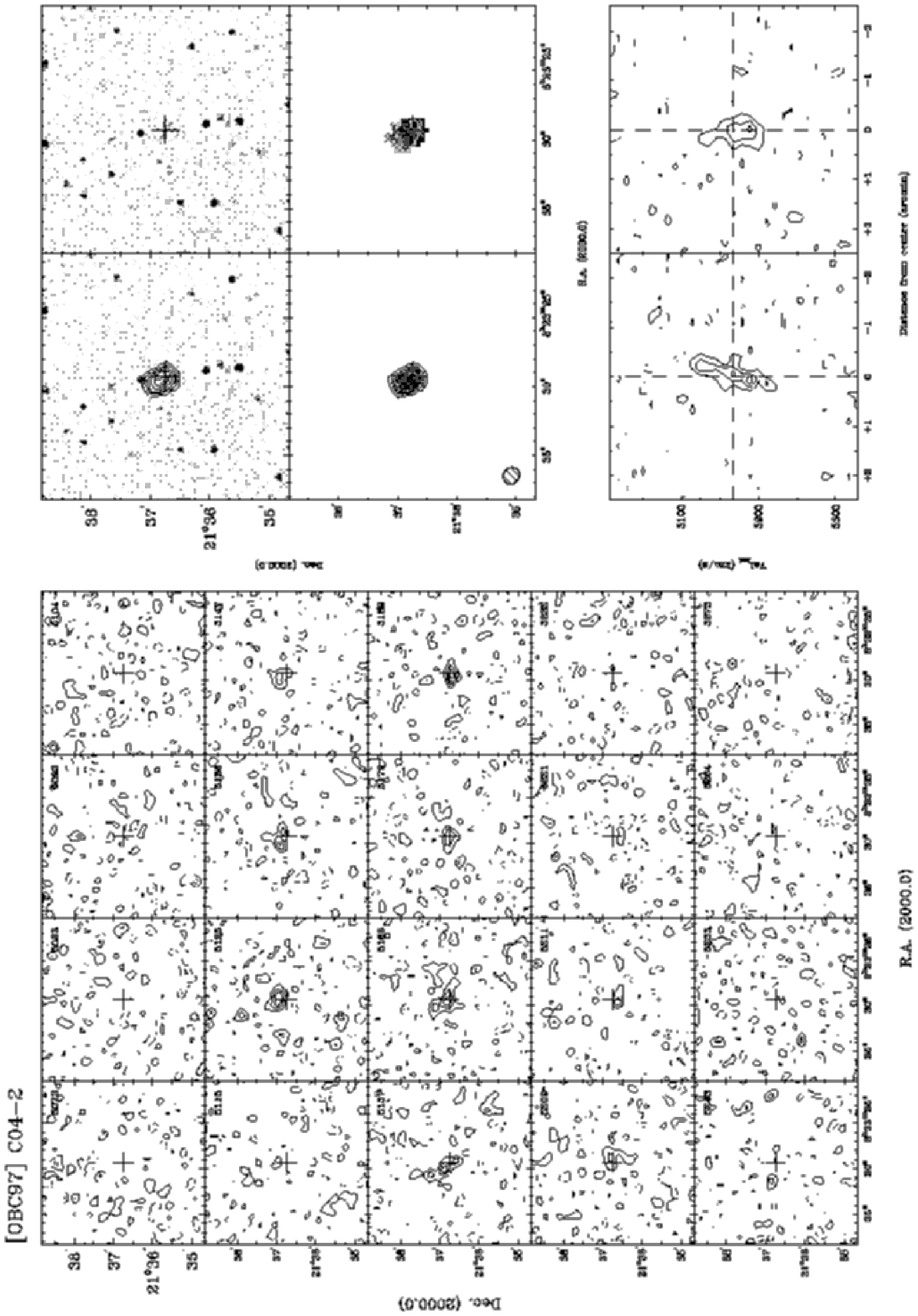}
\caption{\footnotesize
C04-2 ($4.25^{'}\times4.25^{'}$); LSB.
Left) Channel maps. The cross in each box indicates the optical center. 
The velocities in the upper right conners are heliocentric.
Contour levels are 0.47$\times$1, 2, 3... mJy beam$^{-1}$ (solid lines) 
and -0.47$\times$1, 2 mJy beam$^{-1}$ (dashed lines). 
Right-Top) Counterclockwise from the top right, the optical image; 
the HI contours on the optical image (contours are 0.18+0.54$\times$1, 2, 3,...
10$^{19}$ cm$^{-2}$ with the peak of 2.59$\times$10$^{19}$ cm$^{-2}$); 
the HI contours on the grayscales; the velocity field map (5168$\pm$5 km s$^{-1}$). 
Right-Bottom) Position-velocity cuts - along the major axis (left) 
and the minor axis (right). Contour levels are 0.55$\times$1, 2, 3... mJy beam$^{-1}$ 
(solid lines) and -0.55$\times$1, 2 mJy beam$^{-1}$ (dashed lines).\label{fig:fig4}}
\end{figure}

\clearpage
\begin{figure}
\plotone{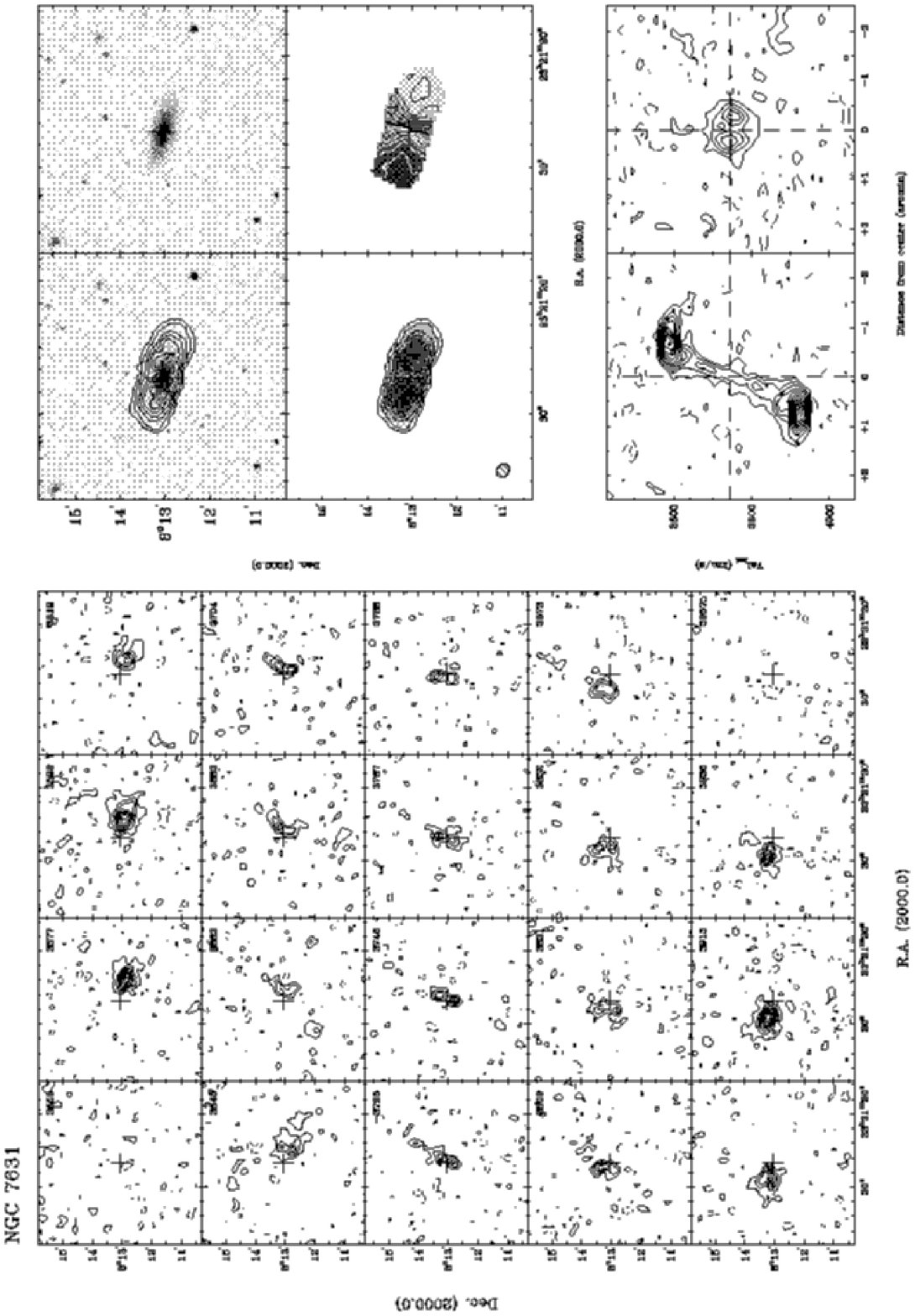}
\caption{\footnotesize
NGC 7631 ($5.6^{'}\times5.6^{'}$). 
Left) Channel maps. The cross in each box indicates the optical center. 
The velocities in the upper right conners are heliocentric. 
Contour levels are 0.73$\times$1, 2, 3... mJy beam$^{-1}$ (solid lines) 
and -0.73$\times$1, 2 mJy beam$^{-1}$ (dashed lines). 
Right-Top) Counterclockwise from the top right, the optical image; 
the HI contours on the optical image (contours are 0.64+1.6$\times$1, 2, 3,...
10$^{19}$ cm$^{-2}$ with the peak of 9.74$\times$10$^{19}$ cm$^{-2}$); 
the HI contours on the grayscales; the velocity field map (3771.9$\pm$25 km s$^{-1}$). 
Right-Bottom) Position-velocity cuts - along the major axis (left) 
and the minor axis (right). Contour levels are 0.56$\times$1, 2, 3... mJy beam$^{-1}$ 
(solid lines) and -0.56$\times$1, 2 mJy beam$^{-1}$ (dashed lines).\label{fig:fig5}}
\end{figure}

\clearpage
\begin{figure}
\plotone{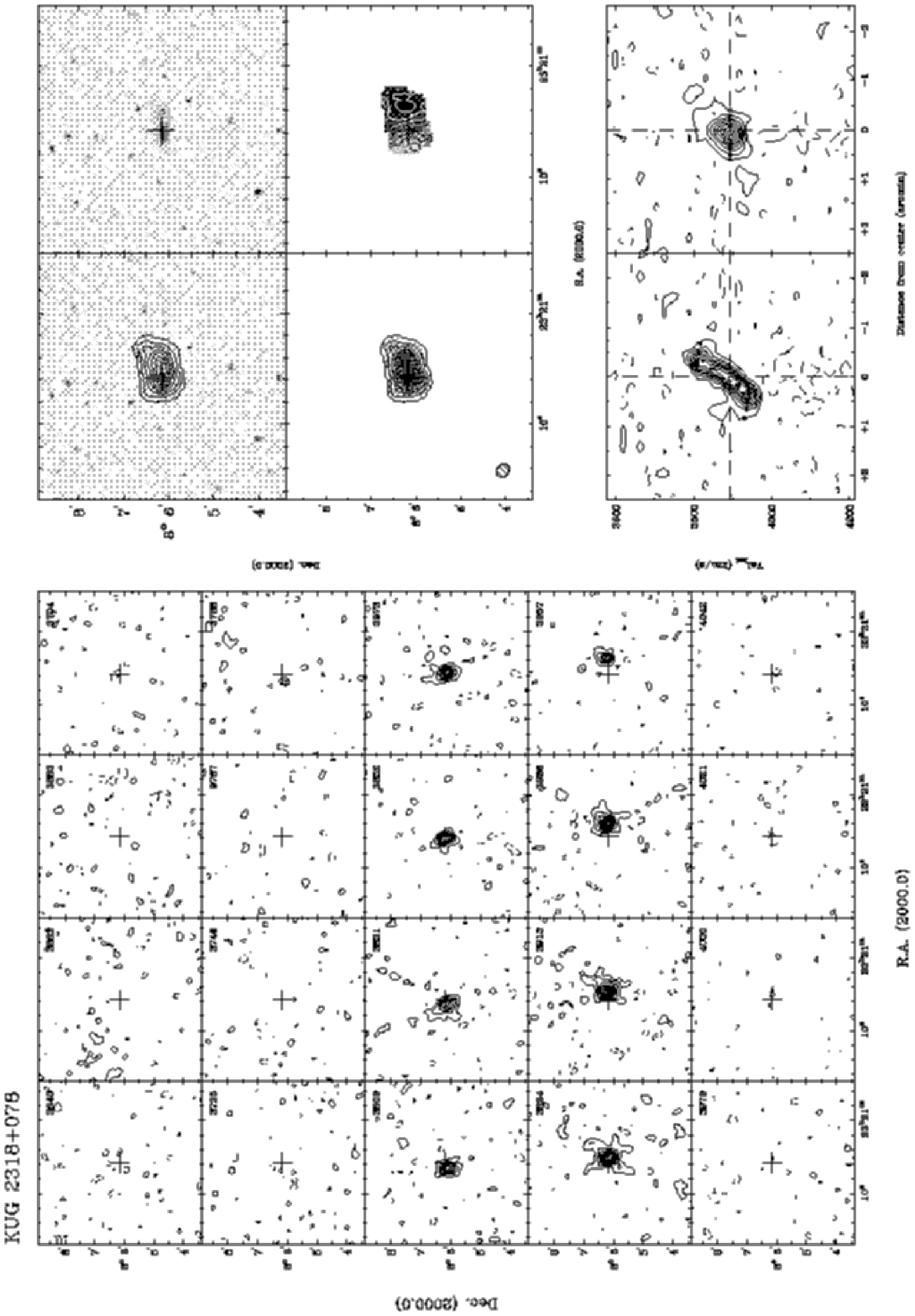}
\caption{\footnotesize
KUG 2318+078($5.6^{'}\times5.6^{'}$). 
Left) Channel maps. The cross in each box indicates the optical center. 
The velocities in the upper right conners are heliocentric. 
Contour levels are 0.78$\times$1, 2, 3... mJy beam$^{-1}$ (solid lines) 
and -0.78$\times$1, 2 mJy beam$^{-1}$ (dashed lines). 
Right-Top) Counterclockwise from the top right, the optical image; 
the HI contours on the optical image (contours are 0.64+1.92$\times$1, 2, 3,...
10$^{19}$ cm$^{-2}$ with the peak of 13.5$\times$10$^{19}$ cm$^{-2}$); 
the HI contours on the grayscales; the velocity field map (3896.1$\pm$15 km s$^{-1}$). 
Right-Bottom) Position-velocity cuts - along the major axis (left) 
and the minor axis (right). Contour levels are 0.54$\times$1, 2, 3... mJy beam$^{-1}$ 
(solid lines) and -0.54$\times$1, 2 mJy beam$^{-1}$ (dashed lines).\label{fig:fig6}}
\end{figure}

\clearpage
\begin{figure}
\plotone{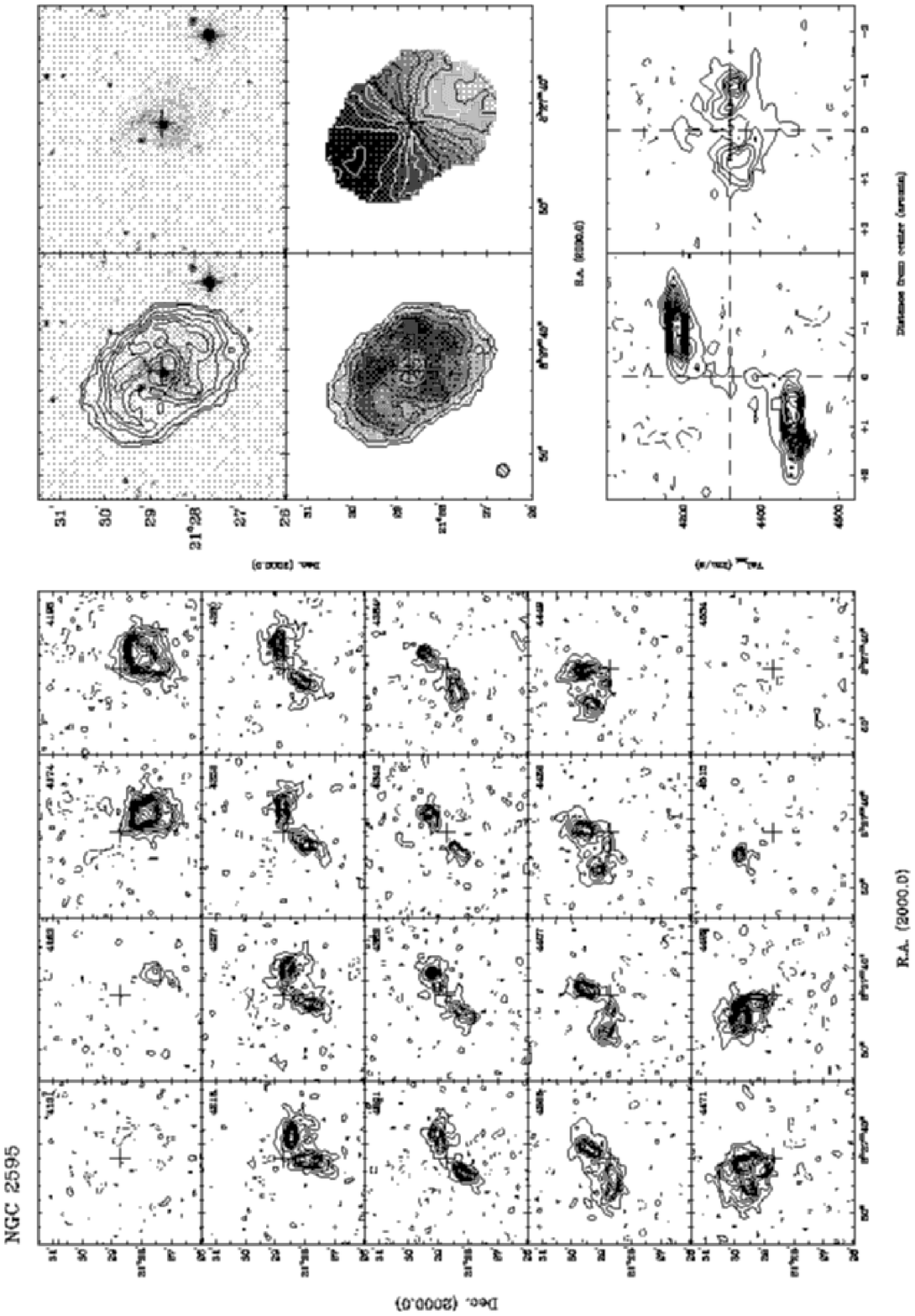}
\caption{\footnotesize
NGC 2595 ($5.6^{'}\times5.6^{'}$). 
Left) Channel maps. The cross in each box indicates the optical center.  
The velocities in the upper right conners are heliocentric.   
Contour levels are 0.6$\times$1, 2, 3... mJy beam$^{-1}$ (solid lines) 
and -0.6$\times$1, 2 mJy beam$^{-1}$ (dashed lines). 
Right-Top) Counterclockwise from the top right, the optical image; 
the HI contours on the optical image (contours are 0.8+1.92$\times$1, 2, 3,...
10$^{19}$ cm$^{-2}$ with the peak of 14.2$\times$10$^{19}$ cm$^{-2}$); 
the HI contours on the grayscales; the velocity field map (4330.8$\pm$25 km s$^{-1}$). 
Right-Bottom) Position-velocity cuts - along the major axis (left) 
and the minor axis (right). Contour levels are 0.58$\times$1, 2, 3... mJy beam$^{-1}$ 
(solid lines) and -0.58$\times$1, 2 mJy beam$^{-1}$ (dashed lines).\label{fig:fig7}}
\end{figure}

\clearpage
\begin{figure}
\plotone{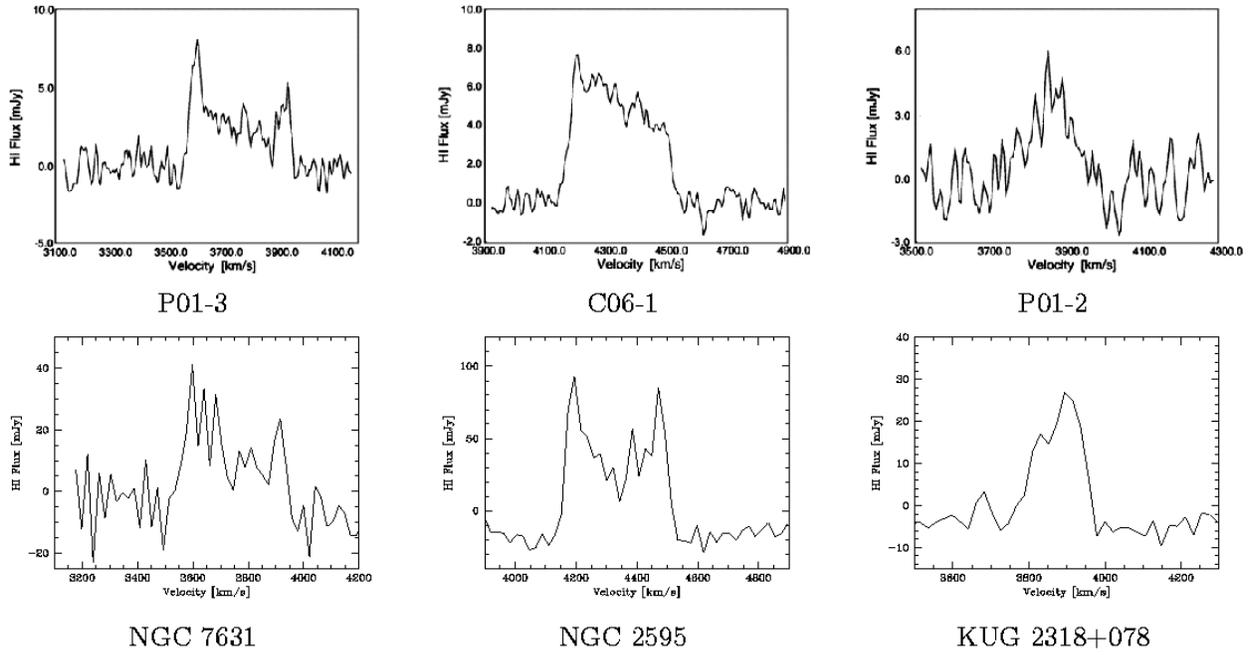}
\caption{\footnotesize
The HI profiles of the extreme non TF LSB galaxies obtained using 
the 305m Arecibo Gregorian Telescope (top; P01-3, C06-1, P01-2) \citep{one00a} 
and the global HI profiles of the bright systems observed with the VLA (bottom; 
NGC 7631, NGC 2595, KUG 2318+078). Note that the HI spectra of each pair of a 
LSB galaxy and its bright companion (P01-3 \& NGC 7631, C06-1 \& NGC 2595, P01-2 
\& KUG 2318+078) is consistent and the distance between the LSB system and its 
bright neighbor is less than 3 arcminute.\label{fig:fig8}}
\end{figure}

\clearpage
\begin{figure}
\plotone{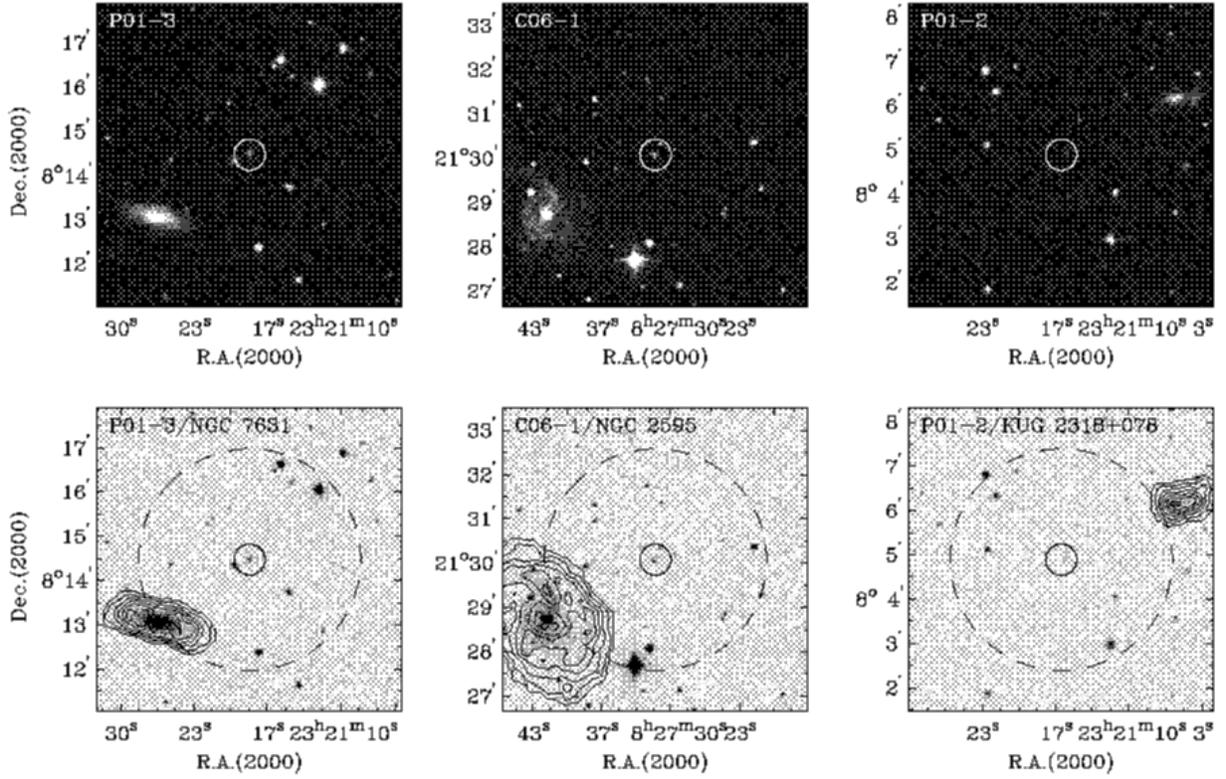}
\caption{\footnotesize
Top) Optical images of P01-3, C06-1 and P01-2 ($7^{'}\times7^{'}$). The 
LSB galaxies are located inside of little circles of 0.7$^{'}$ diameter. Each 
case has a bright companion at the close distance. Bottom) Integrated HI maps 
overlaid on optical images. The outer circle (the dashed line) indicates 5$^{'}$
where the main Arecibo beam reaches 20\% the peak on average at 1.4 GHz. 
Again, the inner circle is the size of 0.7$^{'}$ diameter.\label{fig:fig9}}
\end{figure}

\clearpage
\begin{figure}
\plotone{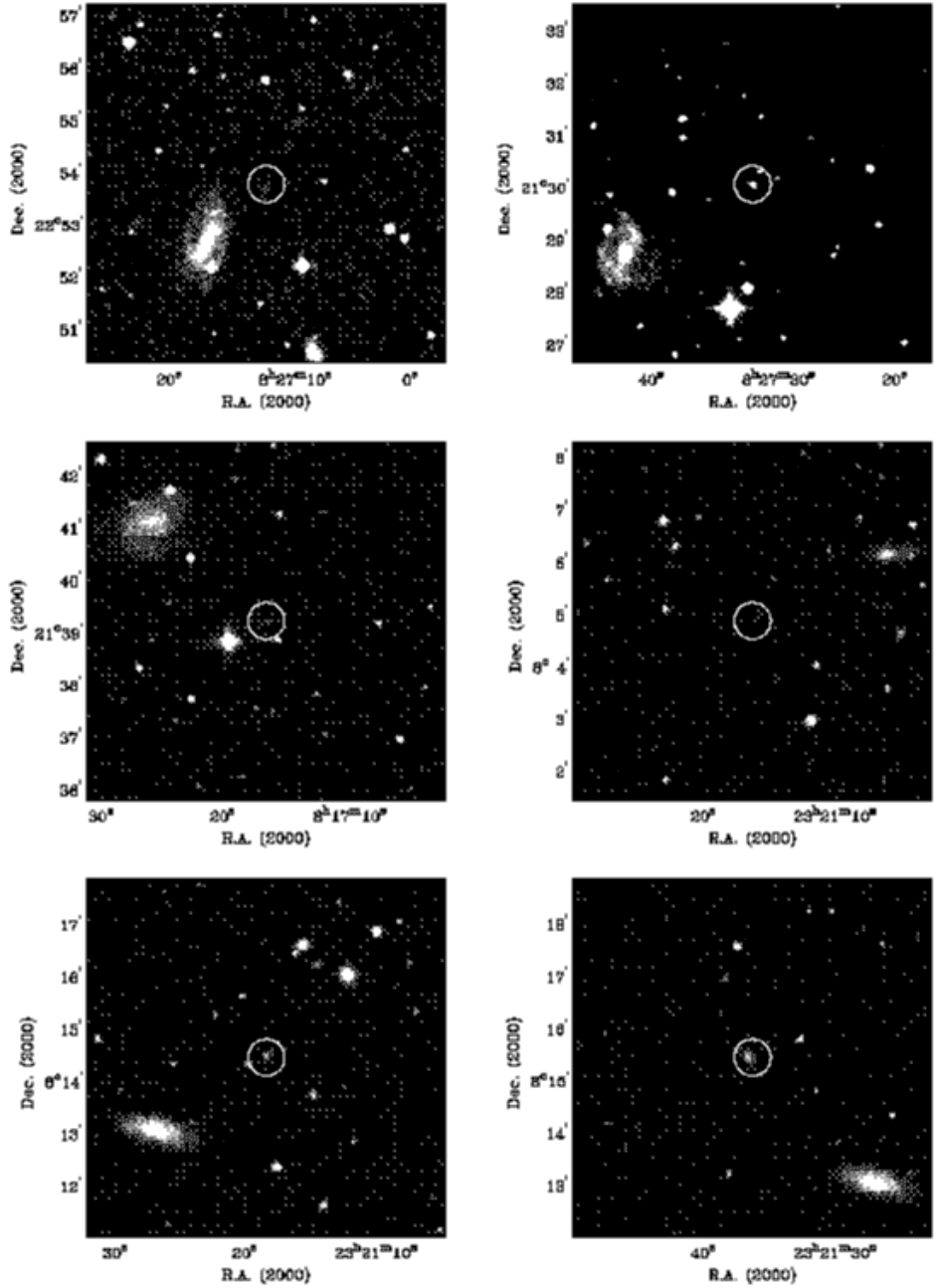}
\caption{\footnotesize
The HI detected LSB galaxies \citep{one00a} which have bright companions 
in the same velocity ranges. The LSB systems are located in the center of panels 
of $7^{'}\times7^{'}$. Top (from left to right)- C05-5 \& UGC 4416 and C06-1 \& NGC 
2595; Middle- C08-3 \& UGC 4308 and P01-2 \& KUG 2318+078; Bottom- P01-3 \& NGC 7631 
and P01-4 \& NGC 7631.\label{fig:fig10}}
\end{figure}

\clearpage
\begin{figure}
\plotone{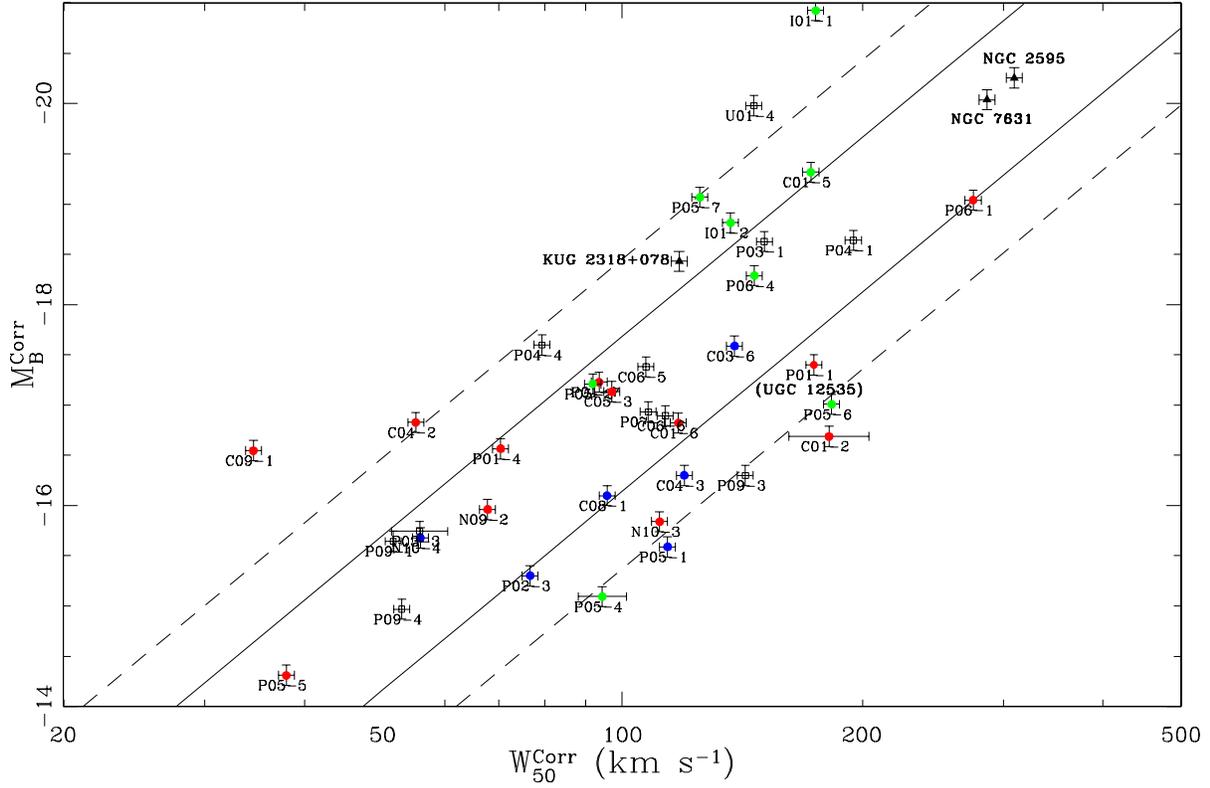}
\caption{\footnotesize
The Tully-Fisher relation for \citet{one00a}'s sample excluding contaminated LSB 
galaxies. Bright galaxies, NGC 7631, NGC 2595, KUG 2318+078 (filled triangle) and 
one LSB galaxy P01-1 (UGC 12535) are added. Note NGC 7631 and NGC 2595 fall into the 
center of the TF. The empty circles indicate the very blue ($B-V<$0.6) LSB galaxies, 
the filled circles indicate very red ($B-V>$0.8) LSBs, and the stars are intermediate 
colors (0.6$\leq B-V \leq$0.8). While the red LSBs (filled circles) are spread in both 
sides of the TF relation, the blue ones (empty circles) tend to be underluminous at a 
given linewidth. If LSBs are {\rm\sc Hi}-detected by \citet{one00a} but their colors 
are not available \citep{one97a}, they are plotted with open squares. The errorbars 
indicate 1$\sigma$ in both axes.\label{fig:fig11}}
\end{figure}

\clearpage
\begin{figure}
\plottwo{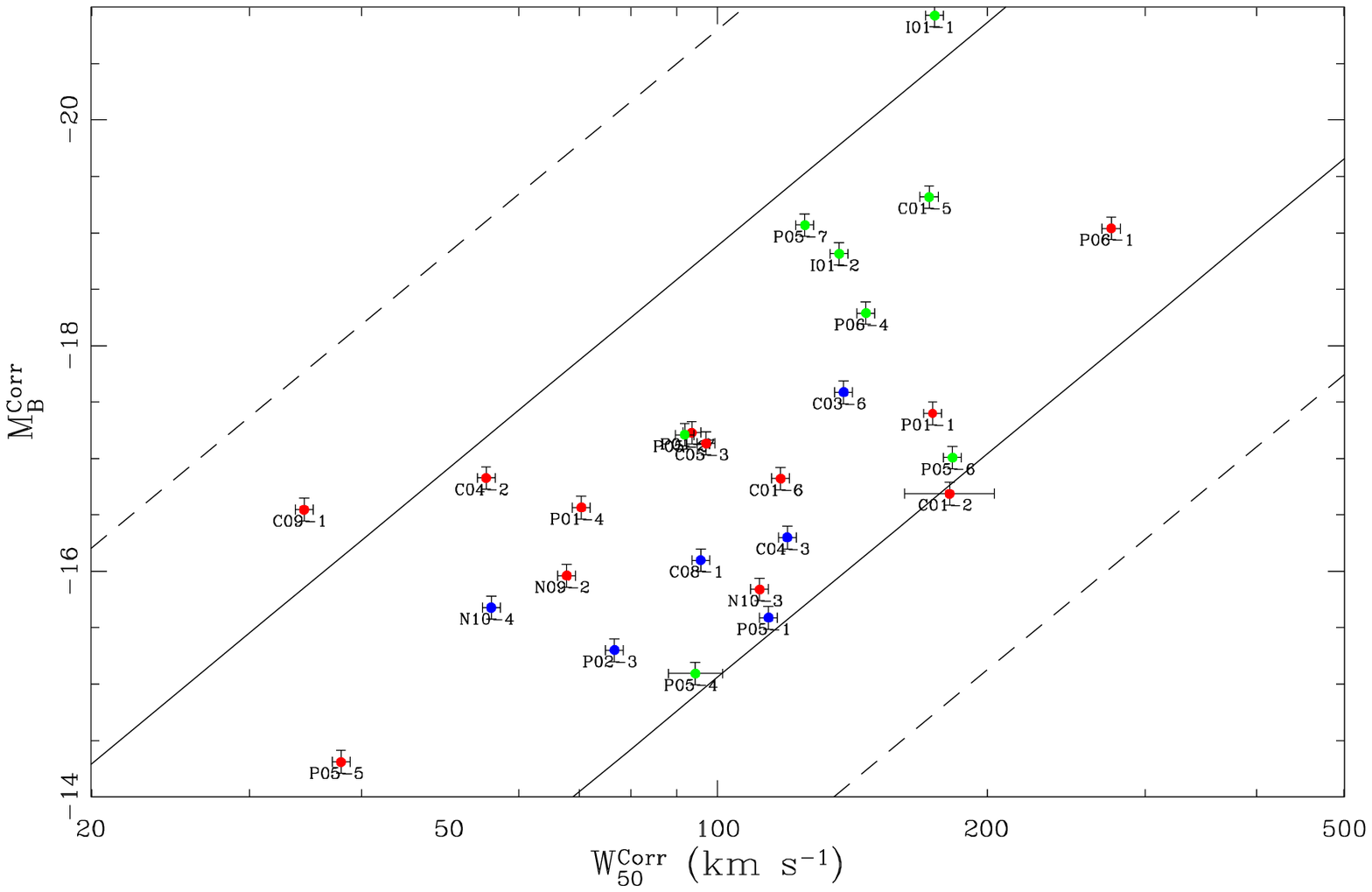}{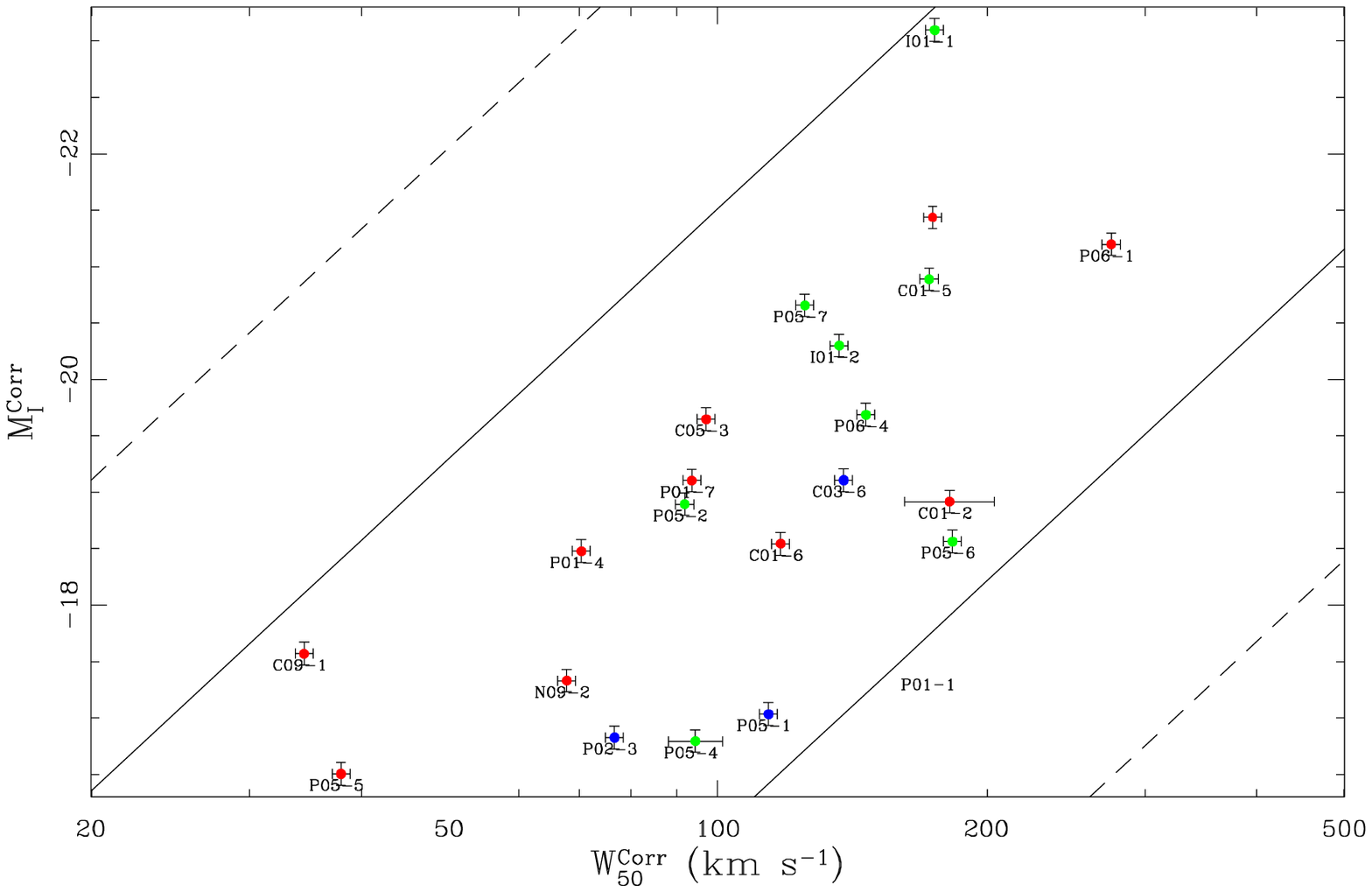}
\caption{\footnotesize
The Tully-Fisher relation: Left) in B-band; 
M$_B^{Corr}$=-6.568$\times$log(W$_{50}^{Corr}$)+(-3.838$\pm$1.913). 
The same plot as Fig. 11, but high surface brightness galaxies are dropped
here. Also, instead of using the TF relation obtained by \citet{zwa95}, 1-$\sigma$ and 
2-$\sigma$ lines are drawn with a newly derived TF relation with \citet{one00a}'s sample.
Right) I-band TF relation; M$_I^{Corr}$=-7.372$\times$log(W$_{50}^{Corr}
$)+(-4.009$\pm$2.757). In $I$-band, it is plotted with only 21 galaxies of which 
magnitudes in $I$-band are available. The color classification is same as
Fig. 11.\label{fig:fig12}}
\end{figure}

\end{document}